\documentclass[conference]{IEEEtran}
\IEEEoverridecommandlockouts
\usepackage{cite}
\usepackage{amsmath,amssymb,amsfonts}
\usepackage{algorithmic}
\usepackage{graphicx}
\usepackage{textcomp}
\usepackage{xcolor}
\usepackage{graphicx}
\usepackage{subcaption}
\def\BibTeX{{\rm B\kern-.05em{\sc i\kern-.025em b}\kern-.08em
    T\kern-.1667em\lower.7ex\hbox{E}\kern-.125emX}}
\begin{document}

\title{Bispectrum Analysis of Noninvasive EEG Signals Discriminates Complex and Natural Grasp Types\\
\thanks{This research was partially supported by the RI-INBRE program ( through NIH P20GM103430) and NSF grant award ID 2245558.}
}

\author{
\authorblockN{Sima Ghafoori, Ali Rabiee, Anna Cetera, Reza Abiri}
\authorblockA{Department of Electrical, Computer and Biomedical Engineering, University of Rhode Island, Kingston, RI, USA \\ Emails: \{sima.ghafoori, ali.rabiee, annacetera, reza\_abiri\}@uri.edu}
 }

\maketitle

\begin{abstract}
The bispectrum stands out as a revolutionary tool in frequency domain analysis, leaping the usual power spectrum by capturing crucial phase information between frequency components.  In our innovative study, we have utilized the bispectrum to analyze and decode complex grasping movements, gathering EEG data from five human subjects. We put this data through its paces with three classifiers, focusing on both magnitude and phase-related features. The results highlight the bispectrum's incredible ability to delve into neural activity and differentiate between various grasping motions with the Support Vector Machine (SVM) classifier emerging as a standout performer. In binary classification, it achieved a remarkable 97\% accuracy in identifying power grasp, and in the more complex multiclass tasks, it maintained an impressive 94.93\% accuracy. This finding not only underscores the bispectrum's analytical strength but also showcases the SVM's exceptional capability in classification, opening new doors in our understanding of movement and neural dynamics.
\end{abstract}


\section{Introduction}
Reaching and grasping movements are imperative yet seemingly straightforward daily activities to which we do not give much thought while performing them \cite{castiello2005neuroscience, errante2021grasping}. Even though prehension is done with ease, its underlying neural activities have been proven to be complex and have been investigated using various research techniques and methods\cite{errante2021decoding,breveglieri2023complementary}. Moreover, studies have taken a path to delve into the concept of grasping activities by deploying more sophisticated analysis methodologies to detect and discriminate different types of movements based on various modalities \cite{volkova2019decoding}. Their aim is not only to study the background neuroscience but also to employ their findings for controlling neuroprosthetic devices using human intentions \cite{volkova2019decoding}. 
Given that the ultimate agenda is presenting a safe, portable, and affordable setup, there has been rising interest in using Electroencephalography (EEG) as a prominent biomarker for such research \cite{bodda2022exploring}. There has been an eclectic mix of techniques, procedures, and platforms offered by articles for acquiring and studying information about different reach-to-grasp movements \cite{xu2021decoding,schwarz2017decoding,schwarz2020analyzing,sburlea2021disentangling}. Furthermore, how they proceed with outlining their experimental setup is diverse. Moreover, the studies have inspected the EEG signals using different temporal and spectral analysis methods. The prominent method for analyzing signals temporally is using Movement-Related Cortical Potential (MRCP) components. In terms of spectral analysis, Power Density functions (PSD) have a significant play alongside the less-used method of wavelet transform. Moreover, few articles employed higher-order analytic methods such as bispectrum to study imagined movements \cite{saikia2011bispectrum, hrisca2021higher}. To the best of our knowledge, no articles looked into bispectrum for discrimination of motor executed reach-to-grasp movements. Further, for distinguishing the defined reach-to-grasp movements in articles, Machine Learning (ML) techniques such as Support Vector Machines, Random Forest (RF), Linear Regression, and the most frequent one, Linear Discriminant Analysis (LDA) have been employed \cite{xu2021decoding,schwarz2017decoding,schwarz2020analyzing,sburlea2021disentangling}.  
Hence, in the current study, we opted for exploiting a state-of-the-art approach using bispectrum to study the underlying neural activities behind two types of the most common daily graspings, precision grip and power grip. Further, we applied ML to distinguish the two movements. Hence, to acquire the informative data, we designed a new platform and used pen and bottle for power and precision grasps respectively. We also incorporated a space for no movement. 

\begin{figure}[!ht]
\centering
\includegraphics[width=3in]{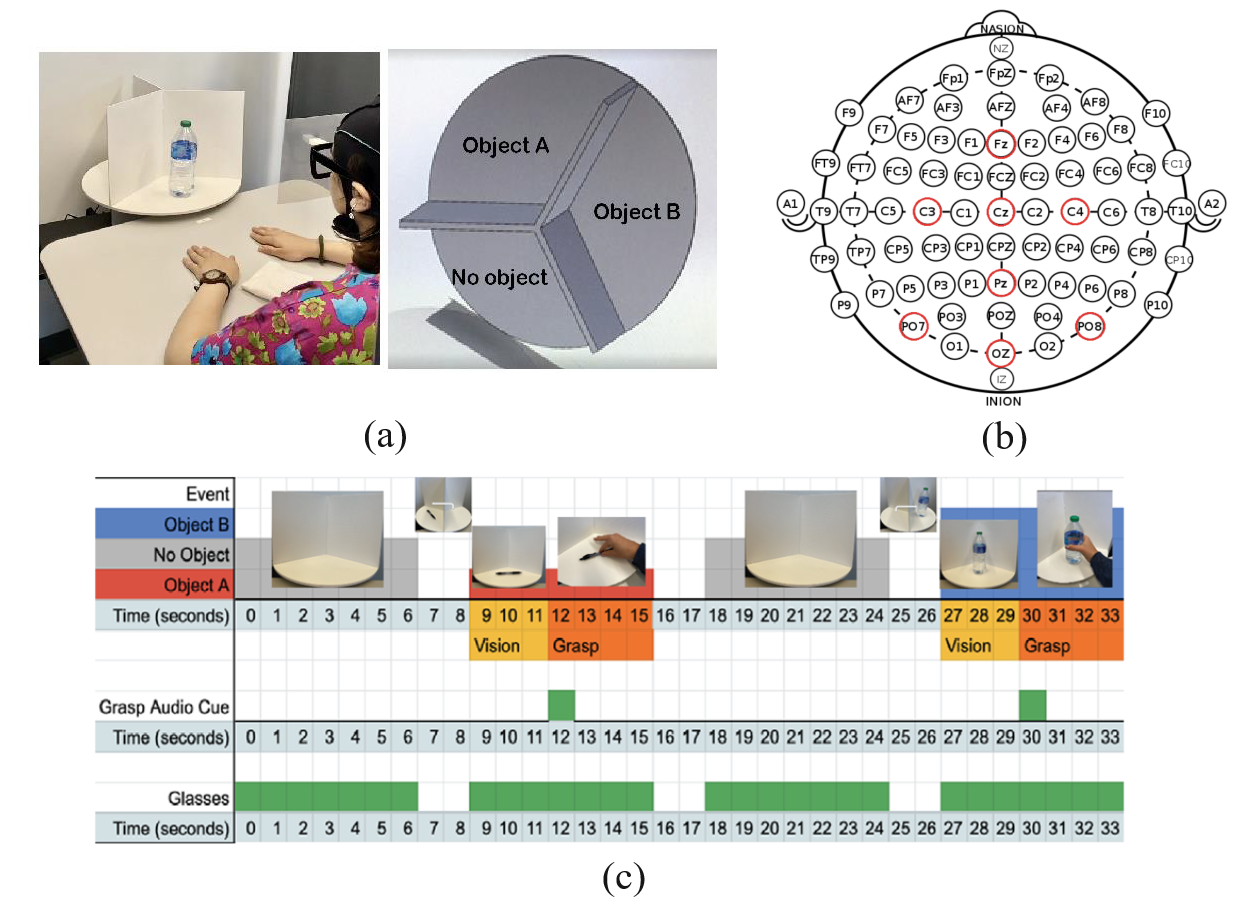}
\caption{Representation of (a) the designed platform, (b) locations of EEG electrodes, and (c) experimental setup}
\label{setup}
\end{figure}

\section{Methods}

\subsection{Platform Design and Experimental Paradigm}
In the experimental setup, a 3D-designed, motorized turntable divided into three sections was used to present two objects (a bottle and a pen) and a no-object scenario. The turntable was operated by a computer, utilizing a TB6600 4A 9-42V Stepper Motor Driver and a Nema 17 Stepper Motor with 1.7A Bipolar. Custom Python software was employed to ensure the synchronized functioning of all hardware and software elements, as well as for logging events, which was critical for processing EEG signals. Participants wore "smart eyeglasses" that were PC-controlled to switch between transparent and opaque states using a smart film. This mechanism was key in controlling the visibility of objects to the participants, ensuring their reactions were unbiased. During the experiment, participants sat in a comfortable chair, maintaining a neutral posture with their palms down and positioned 30 cm from the object's center. As shown in Figure 1c, after observing for 3 seconds, a buzzing sound was emitted as an auditory cue, signaling the start of the grasping phase with their dominant hand, constrained to 3 seconds.  After each grasping phase, the turntable moved on to the next object or the no-object condition. During this transition, the smart eyeglasses were made opaque to avoid any anticipatory bias in the participants. Figure \ref{setup} shows the designed platform and experimental paradigm. 

\subsection{ Data Recording and Preprocessing}
EEG measurements were obtained using the Unicorn Hybrid Black headset, a mobile and wireless EEG tool with dry electrodes. It features eight electrodes and records at a frequency of 250 Hz (more information can be found at [Unicorn Hybrid Black headset](https://www.unicorn-bi.com). The electrode arrangement is shown in Figure 1b. All electrodes used the mastoids as a reference point, employing a linked mastoid reference system. For each experimental condition, we carried out 50 trials per participant for each condition.
Our preprocessing steps included detrending, band-pass filtering (1-40 Hz, order of 5), baseline correction using the signals collected during the rotation of the turntable,  Z scoring for normalizing the data, and independent component Analysis (ICA) for artifact removal.  

\subsection{Participants}
In our study, we enlisted five healthy volunteers from the University of Rhode Island, including two males and three females aged 22 to 36 years, all right-handed. None had a history of neurological disorders. Before starting, each participant received a detailed explanation of the experiment and gave written consent. The research methods were approved by the Institutional Review Board (IRB) to ensure they met ethical research standards.

\subsection{Bispectrum Analysis}\label{AA}
The bispectrum is a higher-order frequency domain analysis tool that is particularly useful in signal processing for identifying phase coupling and non-linear interactions in signals. It is an extension of the power spectrum, a second-order statistic, to the third-order. For a continuous signal \( x(t) \), the bispectrum \( B(f_1, f_2) \) is defined as the triple product of the Fourier transform of the signal at different frequencies. Mathematically, it can be expressed as \cite{saikia2011bispectrum}:

\[ B(f_1, f_2) = \int_{-\infty}^{\infty} \int_{-\infty}^{\infty} X(f_1) X(f_2) X^*(f_1 + f_2) \, df_1 \, df_2 \]

where \( X(f) \) is the Fourier transform of \( x(t) \), and \( X^*(f) \) is the complex conjugate of \( X(f) \), and \( f_1 \) and \( f_2 \) are frequency components. In this study, Bispectrum was calculated using a 256-point Fast Fourier Transform (FFT) with Hanning windows and 50 percent overlap for time segments. 

\subsection{Feature Extraction and classification}
We extracted a comprehensive set of features from the bispectrum of the whole period of observation and execuation of each trial. It included both amplitude-based and phase-related features. For the amplitude-based features, the average bispectral power, the maximum bispectral power, the sum of absolute bispectral power, the bispectral entropy, and the maximum bicoherence we derived. These metrics collectively provided a detailed quantitative assessment of the bispectrum's power distribution and complexity.

In addition to these, we also delved into phase-related features, which were crucial for understanding the phase relationships in the bispectrum. Therefore, we computed the bispectral phase coherence in its real form, the mean phase angle, and the second moment of the phase, offering insights into the uniformity and variability of the phase data. Furthermore, phase-related entropy and the real part of phase bicoherence were also added, adjusting the calculation to account for normalization and providing a more nuanced understanding of phase information. That left us with 80 features for each trial (10 features from each channel).  Using a 5-fold stratified cross-validation approach, three classifiers (Random Forst: RF, Support Vector Machin: SVM, Linear Discriminant Analysis: LDA) were trained and tested on our dataset. It is worth noting that the training and testing sets were standardized separately before the development of the models. 

\section{Result}

\begin{figure*}[ht!]
\centering
\begin{subfigure}{.33\linewidth}
\centering
\includegraphics[width=\linewidth]{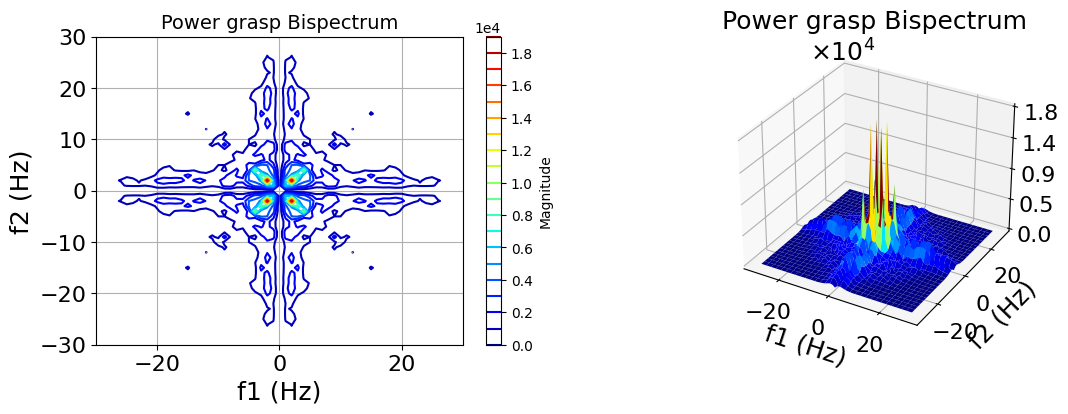}
\end{subfigure}%
\begin{subfigure}{.33\linewidth}
\centering
\includegraphics[width=\linewidth]{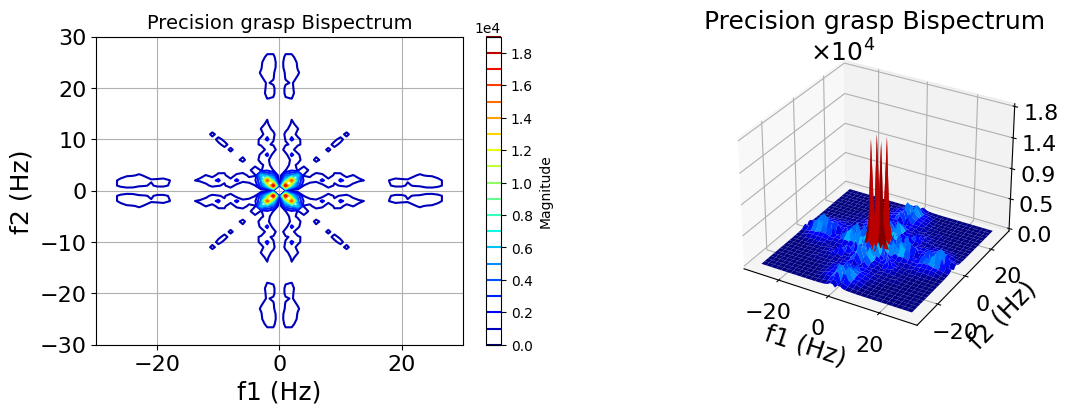}
\end{subfigure}%
\begin{subfigure}{.33\linewidth}
\centering
\includegraphics[width=\linewidth]{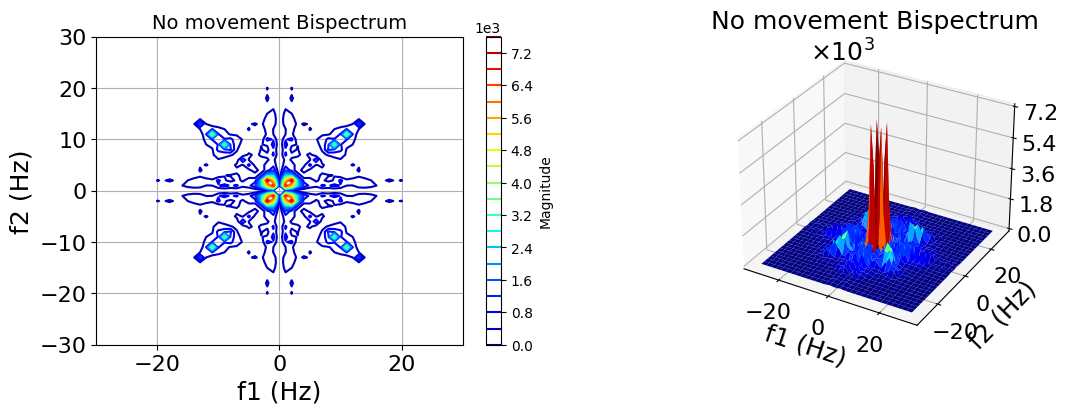}
\end{subfigure}

\begin{subfigure}{.33\linewidth}
\centering
\includegraphics[width=\linewidth]{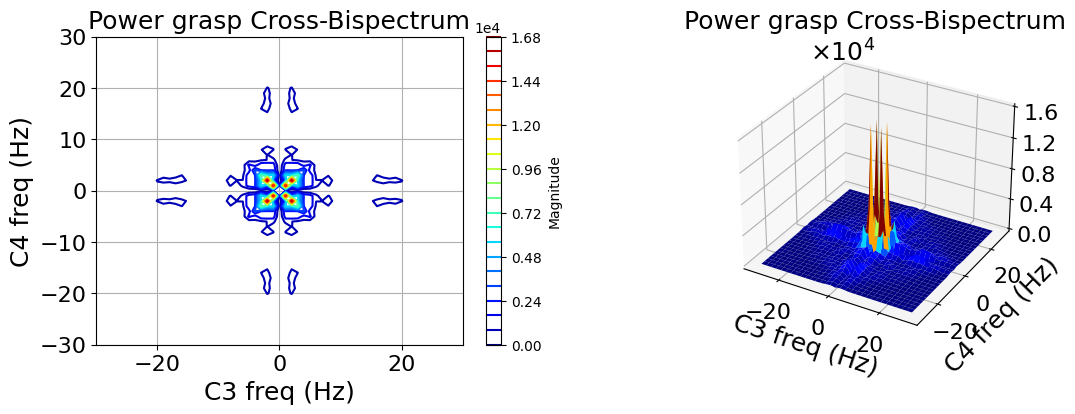}
\end{subfigure}%
\begin{subfigure}{.33\linewidth}
\centering
\includegraphics[width=\linewidth]{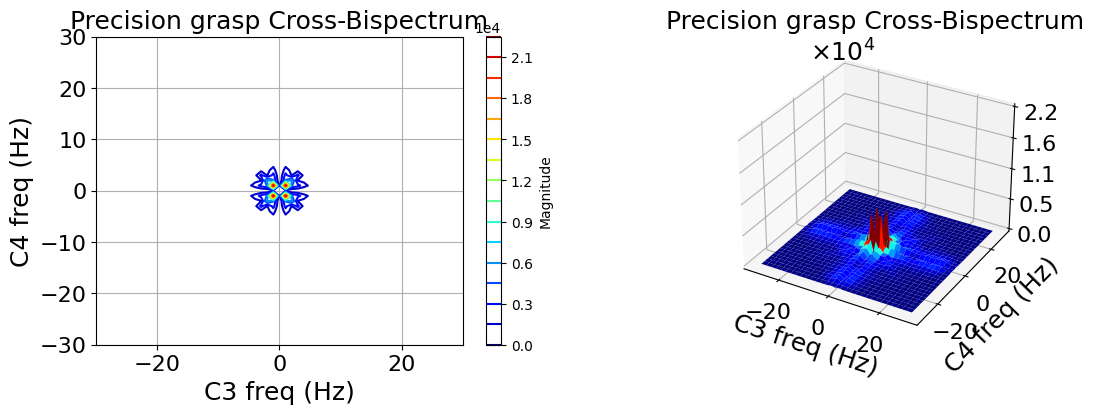}
\end{subfigure}%
\begin{subfigure}{.33\linewidth}
\centering
\includegraphics[width=\linewidth]{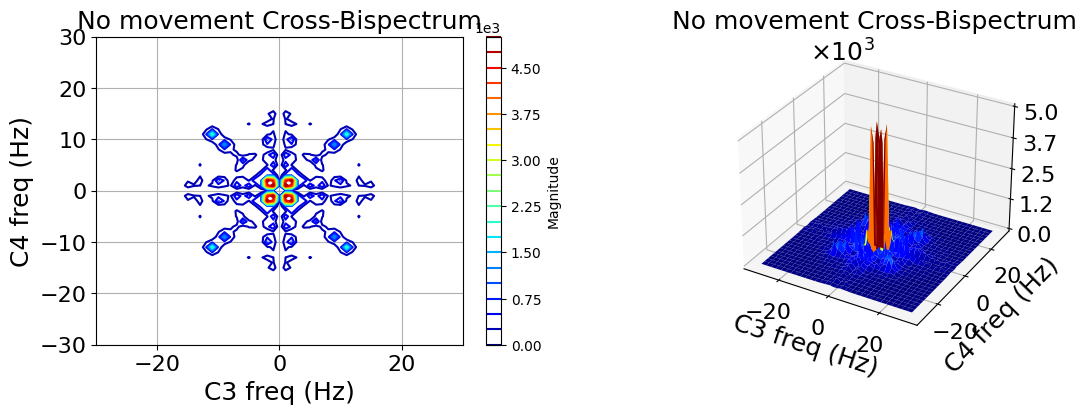}
\end{subfigure}
\caption{Average of bispectrum (top row) contour and 3D plots for C3 channel and cross-bispectrum (bottom row) magnitude plots for channels C3 and C4}
\label{combined}
\end{figure*}

\begin{table*}[h]
\caption{5-fold cross-validated accuracy scores for multiclass and binary classifications, reported for the 5 subjects (\%).}
\centering
\renewcommand{\arraystretch}{1.5}
\begin{tabular}{*{13}{p{0.052\textwidth}}} \hline
 &\multicolumn{3}{c}{Multiclass}&\multicolumn{3}{c}{Power Grasp vs No movement} & \multicolumn{3}{c}{Precision Grasp vs No movement}& \multicolumn{3}{c}{Power Grasp vs Precision Grasp}\\
 \cline{2-13}
 &  RF& SVM&LDA&  RF& SVM&LDA & RF& SVM& LDA& RF& SVM&LDA\\ 
 \cline{2-13}
 S1 & 93.78  & 95.56 & 79.56 & 96.67 & 96.33 &  88.67 & 92.67 & 95.67  & 86.67  & 68.4  &  73 & 65.2 \\ 
 S2 &  93.55 &  93.12 &  81.11 & 96.33 &  96.67 & 94 &  93.33 &  94 & 86.33 & 72.33 & 69 &  67.08\\ 
 S3 & 92.22 & 95.33 &  71.78 & 94.67  &  96.67 &  85 & 94 & 95.67  &  84.33 & 76 &  75 & 74\\  
 S4 & 94.67 &  94.89 &  81.78 &  95.67 &  98.67 &  92 & 95 & 96.11 &  85 &  71 & 68.86 & 69\\  
 S5 & 97.33 &  95.78 & 82.44 & 99 & 96.67 & 96.33 &  98.67 &  96.67 & 98 & 76.03 &  73.99 & 76 \\  
 Mean & 94.31 & 94.93 & 79.33 & 96.47 & 97 & 91.2 & 94.73 & 95.62 & 88.06 & 72.75 & 71.97& 70.25\\ \hline
 
\end{tabular}
\label{table:accuracy}
\end{table*}
\renewcommand{\arraystretch}{1.5} 

Channels C3 and C4 are pivotal in capturing important information in motor tasks \cite{saikia2011bispectrum, schwarz2020analyzing}. Therefore, the bispectrum of all trials was pooled together and averaged for Channel C3 in three conditions. Figure \ref{combined} (top row) depicts the contour and 3D plots of averaged bispectrum magnitudes for power, precision, and no movements for frequencies less than 30 Hz for channel C3.  The associated images show relationships for both positive and negative frequencies. The bottom row of images also represents the same concept but between two channels of C3 and C4 called cross-bispectrum, which demonstrates the frequency interaction between the two channels over the motor cortex. The bispectrum magnitudes of the two types of movements are about 10 times more than those related to the no-movement tasks. Also, the cross-bispectrum magnitude plots seem to be more focused around certain frequencies less than 20 Hz. 

\begin{figure*}[!ht]
\centering
\begin{subfigure}{.3\linewidth}
\centering
\includegraphics[width=\linewidth]{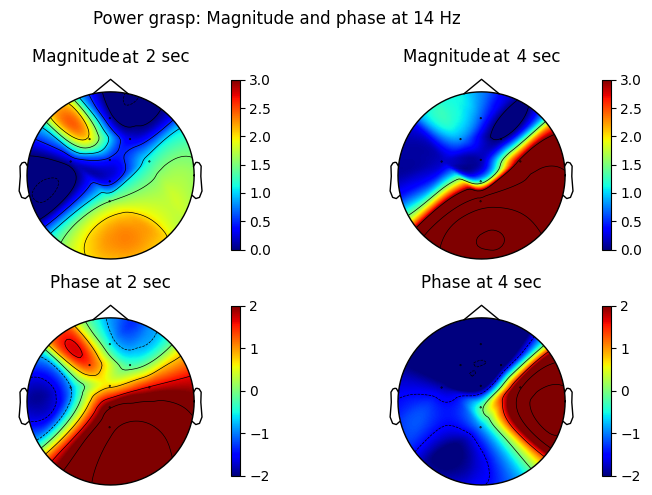}
\end{subfigure}%
\begin{subfigure}{.3\linewidth}
\centering
\includegraphics[width=\linewidth]{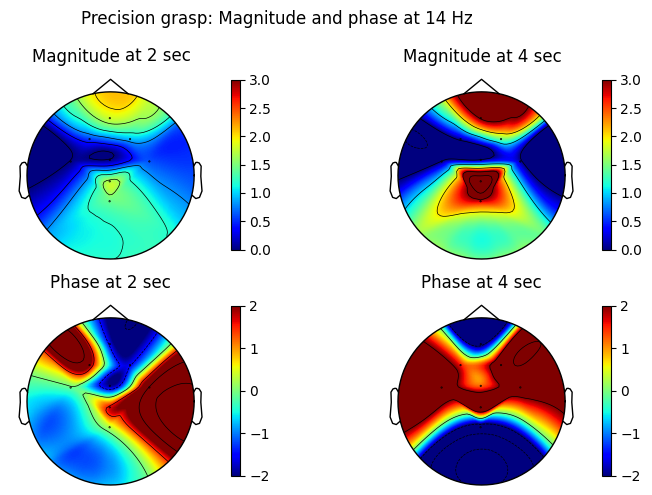}
\end{subfigure}%
\begin{subfigure}{.3\linewidth}
\centering
\includegraphics[width=\linewidth]{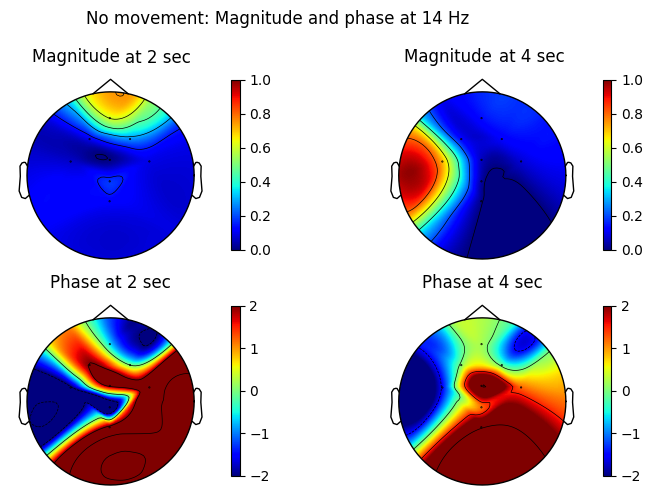}
\end{subfigure}
\caption{Topography plots of Bispectrum Magnitude (top row) and phase (bottom row) at 14 Hz and two-time points during Observation (2nd sec) and grasping (4th sec) in three conditions}
\label{topo}
\end{figure*}

Table \ref{table:accuracy} shows the average accuracy scores of 5-fold cross-validation of the three classifiers for the five subjects using all channels. The accuracies were calculated in 4 classification categories from multiclass to three conditions of binary classification. The power grasp seems to be distinguishable more easily than the precision grasp since the scores obtained from classifiers are notably more in classifying power grip compared to the precision grip in binary classifications.  Also, the data from subject 5 resulted in better results. Overall, SVM outperformed the other two classifiers in most cases with an accuracy of 97\% in power grip versus no-movement classification and 94.93\% in multiclass classification. Notably, the models underperformed in power versus precision classification compared to multiclass classification, with scores around 70\%. 
Figure \ref{topo} shows topography plots of Bispectrum Magnitude and phase at 14 Hz and two-time points, 2nd and 4th seconds, before and after auditory cue in three conditions. During observation, we see higher values for bispectrum magnitude in premotor areas, and during grasping we mostly see heightened activities in motor and occipital areas. The same thing goes for phase interactions. The figure indicates how neural interactions evolve from observation to execution at the frequency of 14 Hz which is within the beta band. 

\section{Discussion}

In this study, we investigated the use of the bispectrum to distinguish between two types of hand movements: power grasping and precision grasping. By analyzing the EEG signals from the C3 channel and between the C3 and C4 channels, we observed distinct patterns in the frequency components of these movements. During power and precision grasping, the bispectrum plots revealed high-magnitude peaks, indicating strong non-linear interactions in the EEG signals. These peaks were more spread out in power grasping, showing a range of frequency interactions, while they were narrower and more defined in precision grasping, suggesting specific frequency interactions during controlled movements. In contrast, the no movement condition showed much lower peaks, indicating less non-linear activity when the hand is at rest. These findings highlight the brain's adaptability and the varying complexities of its electrical activity depending on the motor task. 

Additionally, cross-bispectral analysis revealed phase coupling between signals at different electrodes, particularly between C3 and C4 channels. This analysis showed that power grasp involves a wider neural network, as seen in the high-magnitude peaks, whereas precision grasp involves fewer but more synchronized neural pathways. 
Furthermore, in both the bispectrum and cross-bispectrum plots, power and precision grasps showed concentrated peaks, with precision grasp displaying narrower peaks. The no-movement condition had significantly reduced magnitudes, especially in the cross-bispectrum plots, indicating minimal phase coupling due to lack of motor activity.
Moreover, in terms of classification, generally, RF and SVM performed better than LDA across tasks, with SVM showing the highest mean accuracy in both multiclass and binary classifications. The binary classification between power grasp and no movement achieved the highest mean accuracy, suggesting that this distinction is clearer in the EEG data. Overall, SVM results surpassed the rest in most cases with a mean accuracy of 94.93\% in multiclass, 97\% in power grip vs no-movement,  95.62\% in precision grip vs no-movement, and 71.97\% in power grip vs precision grip classification. RF results were close to SVM and LDA underperformed noticeably compared to the other two classifiers.

The machine learning methods applied in this study are widely recognized and validated in numerous other research projects, and our results significantly surpass the chance level of 33.33\%, demonstrating robust and promising outcomes. The power vs precision task has the lowest mean accuracy, indicating that these two conditions are the hardest to differentiate for the classifiers. These results can inform the development of more accurate models for interpreting EEG data in applications such as brain-computer interfaces.
However, when comparing to the related work, the studies are different not only in their analytic method of choice but also in the number of channels (more than 32 electrodes), grip types, trials, and platform design \cite{xu2021decoding,schwarz2017decoding,schwarz2020analyzing,sburlea2021disentangling}. To the best of our knowledge, no study has ever investigated motor-executed tasks with bispectrum. Studies mostly aim for imaginary tasks \cite{saikia2011bispectrum, hrisca2021higher} which require less sophisticated setups and platforms. Designing a platform like this, which allows for ergonomic movements, presents a significant challenge.

Lastly, the bispectral maps showed that observing an object and grasping it activates different neural networks. Observation mainly activates visual and attentional networks, while grasping also engages motor and sensorimotor networks \cite{castiello2005neuroscience}. This is evident in the stronger magnitude values and specific phase relationships in frequencies related to motor control during grasping, compared to observation. This suggests different neural mechanisms are at play during these tasks, although further analysis is needed to fully understand these activities and interactions.   

\subsection{Limitations and Future Work}

Bispectrum is a higher-order analysis method with so much potential for exploration. For further research, it is recommended to delve into different features and their contribution to classification tasks, to probe phase information of the bispectrum more deeply. We just plot the cross-bispectrum for two specific channels, there is definitely more room to look into that. Also, the phase interactions between frequency bands can be studied in such areas using bispectrum. It is worth noting that the number of channels and trials was limited here. Also, there are other ML models worth evaluating specifically in order to reach better results in discrimination between power and precision movements.

\section{Conclusion}
In our study, we focused on using the bispectrum method to understand brain activity during reach-to-grasp movements. We found that the bispectrum gave us important new insights into how the brain works during these tasks. The SVM model, which used data from the bispectrum, was particularly good at discriminating different kinds of movements.

\bibliographystyle{IEEEtran}
\bibliography{main}

\end{document}